\documentclass[prl,aps,twocolumn,superscriptaddress,showpacs]{revtex4}
\usepackage{amsmath}
\usepackage{amsbsy}
\usepackage{amssymb}
\usepackage[dvips]{graphicx}
\usepackage{epsfig}         

\newcommand{\e}{\epsilon}

\begin{document}

\title{Landscape of solutions in constraint satisfaction problems}

\author{Marc M\'ezard}
\affiliation{Laboratoire de Physique Th\'eorique et
Mod\`eles Statistiques, 
Universit\'e Paris-Sud, F--91405
Orsay, France.}

\author{Matteo Palassini}
\affiliation{Laboratoire de Physique Th\'eorique et
Mod\`eles Statistiques, 
Universit\'e Paris-Sud, F--91405
Orsay, France.}
\affiliation{Departament de F\'\i sica Fonamental,
Universitat de Barcelona, Diagonal 647, E--08028 Barcelona, Spain.}

\author{Olivier Rivoire}
\affiliation{Laboratoire de Physique Th\'eorique et
Mod\`eles Statistiques, 
Universit\'e Paris-Sud, F--91405
Orsay, France.}

\date{\today}

\begin{abstract}
We present a theoretical framework for characterizing the geometrical
properties of the space of solutions in constraint satisfaction problems,
together with  practical algorithms for studying this structure on particular
instances. We apply our method to the coloring problem, for which we obtain
the total number of solutions and analyze in detail the distribution of
distances between solutions.
\end{abstract}


\maketitle

Constraint satisfaction problems (CSPs) offer a unified language describing
many complex systems. Originally investigated by computer scientists in
relation with algorithmic complexity~\cite{Papadimitriou94}, CSPs have
recently attracted much interest within the physics community, following the
discovery of their close ties with spin-glass
theory~\cite{MonassonZecchina99,MezardParisi02}. They are currently used to
tackle systems as diverse as, among others, error-correcting
codes~\cite{BattagliaBraunstein04}, rigidity models~\cite{BarreBishop05a}, and
regulatory genetic networks~\cite{CorrealeLeone04}. The ubiquity of CSPs stems
from their general nature: given a set of $N$ discrete variables subject to
$M$ constraints, a CSP consists in deciding whether there are assignments of
the variables satisfying all the constraints. Of special interest is the class
of NP-complete problems~\cite{Papadimitriou94}, for which no algorithm is
known that guarantees to decide the satisfiability of a problem instance in a
time polynomial in $N$. A well-studied example is the $q$-coloring problem
($q$-COL): given a graph with $N$ nodes and $M$ edges connecting certain pairs
of nodes, and given $q$ colors, can we color the nodes so that no two
connected nodes share a common color?

Much insight into CSPs is gained by focusing on {\em typical}\/ instances
drawn from an ensemble with a fixed density of constraints
$\alpha=M/N$. As $\alpha$ is varied, a threshold phenomenon is generically
observed. Below a critical value $\alpha_c$, instances are typically
satisfiable (SAT phase): at least one satisfying assignment (or {\em solution})
exists with probability one when $N\to \infty$; above $\alpha_c$ they are
typically unsatisfiable (UNSAT
phase). Rigorous bounds on $\alpha_c$ have been derived \cite{AchlioptasNaor05}.
The running time of algorithms often
increases greatly near $\alpha_c$~\cite{CheesemanKanefsky91}.

\begin{figure}
  \begin{minipage}[l]{.58\linewidth}
    \centering \epsfig{file=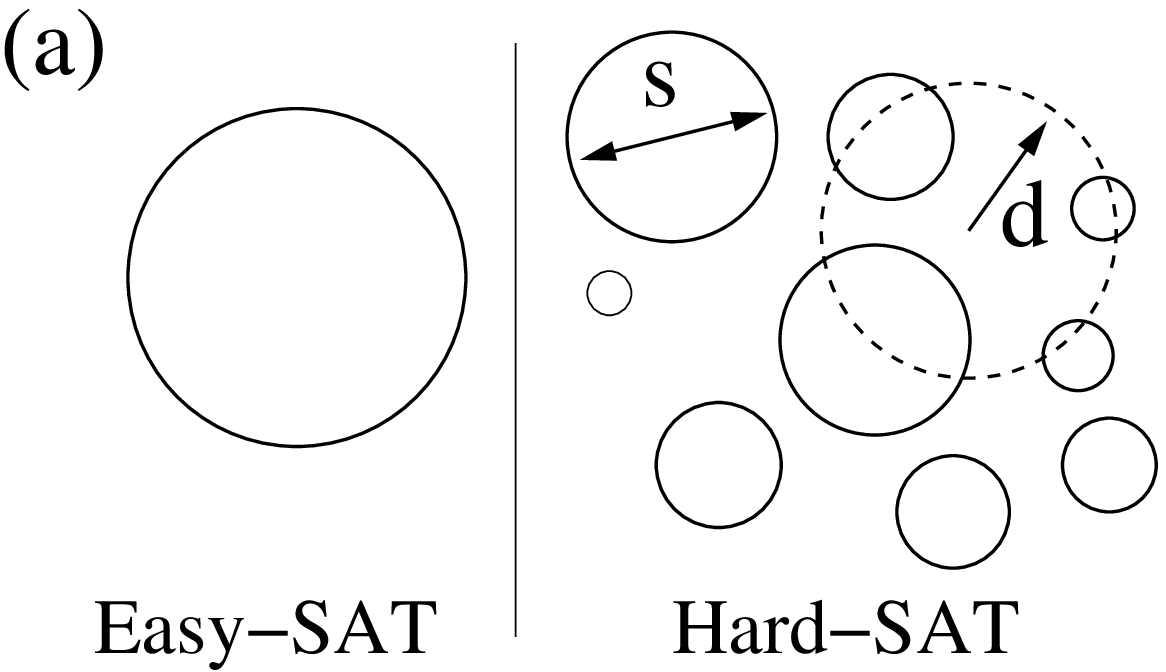,width=4.5cm}
  \end{minipage}
  \begin{minipage}[r]{.40\linewidth}
    \centering \epsfig{file=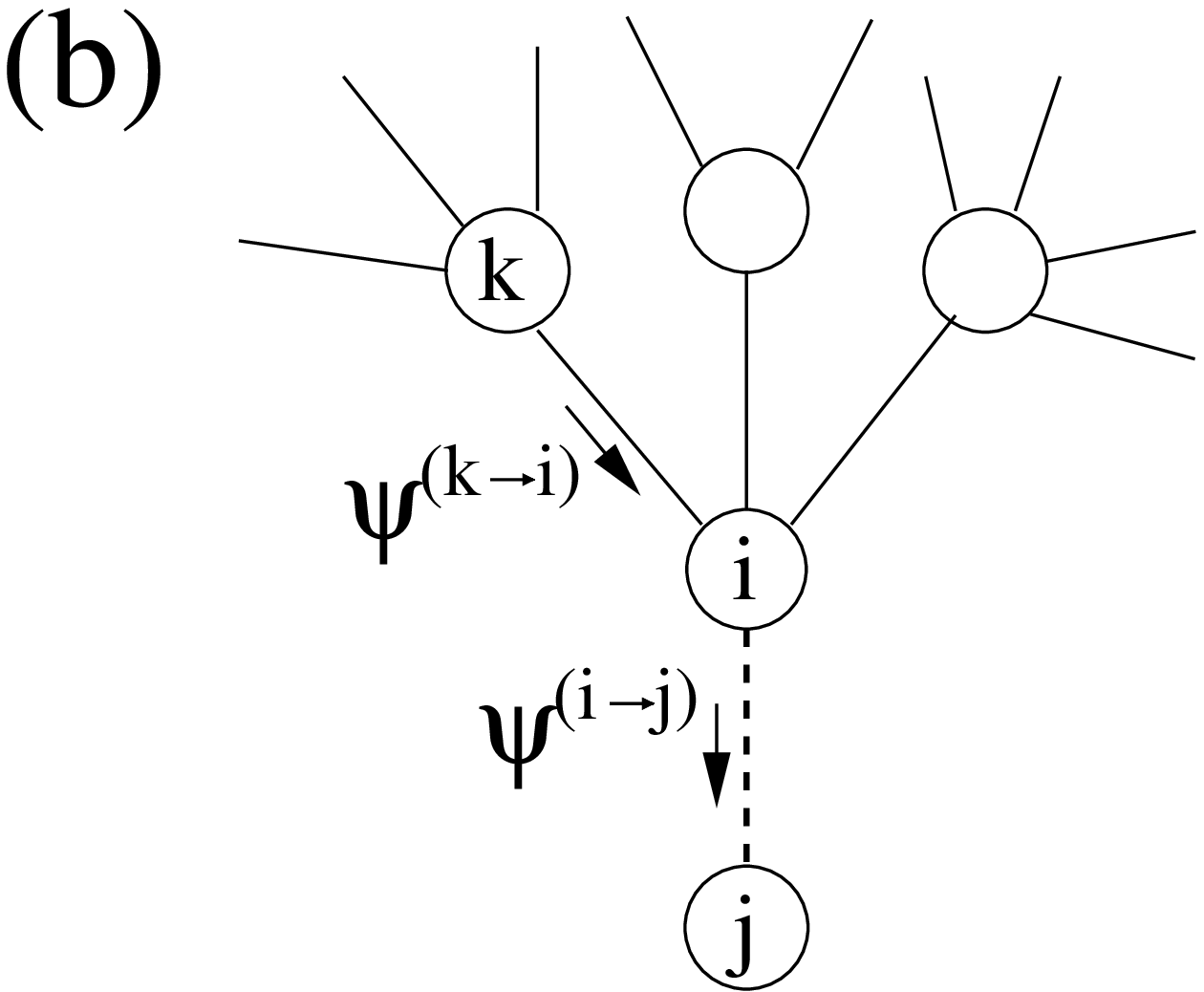,width=3cm}
  \end{minipage}
\caption{\small (a) Illustration of the clustering phenomenon. In the Easy-SAT
phase $\alpha < \alpha_d$, all the solutions are connected. In the
Hard-SAT phase $\alpha_d < \alpha < \alpha_c$, solutions separate into
distinct clusters. (b) Notations used in the cavity approach: the message
(cavity field) $\psi_\sigma^{(i\to j)}$ gives the probability that node $i$
has color $\sigma$ in the absence of node $j$.}\label{fig:scheme}
\end{figure}

CSPs enter the standard framework of statistical physics by associating to
each assignment of the $N$ variables $\sigma\equiv\{\sigma_i\}_{i=1}^N$ an
energy $E[\sigma]$ defined as the number of  constraints violated by $\sigma$. The
satisfiability problem reduces to the determination of the ground-state energy
$E_0=\min_\sigma E[\sigma]$: if $E_0>0$ the instance is UNSAT, if $E_0=0$ it
is SAT. In recent years, several methods borrowed from statistical
physics~\cite{BiroliMonasson00,MezardParisi02,MezardZecchina02} have pointed
to the existence of a {\em second}\/ threshold $\alpha_d<\alpha_c$, associated
with {\em clustering}\/ of the space $\mathcal{S}$ of all solutions. For
$\alpha<\alpha_d$ (Easy-SAT phase), $\mathcal{S}$ is typically connected:  any
two solutions are joined by a path of moves involving a finite number of
variables. For $\alpha_d<\alpha<\alpha_c$ (Hard-SAT phase),
$\mathcal{S}$ is typically disconnected: solutions gather into clusters far
apart from each other (as illustrated in Fig.~\ref{fig:scheme}a), which can
only be joined by moves involving a finite {\em fraction}\/ of the variables.
This scenario, which has been confirmed rigorously in some
cases~\cite{MezardMora05}, suggests that computational hardness 
may be caused by the trapping of local algorithms in {\em metastable}\/ clusters, 
which are exponentially more numerous than clusters of solutions.

In this Letter, we introduce methods to analyze in detail the structure of the
solution space of CSPs in the Hard-SAT phase. The first aspect we analyze is
the entropic structure. A cluster $\lambda$ typically contains an exponential
number of solutions, $\mathcal{M}_\lambda \asymp \exp(N s_\lambda)$, where
$N s_\lambda$ is the internal entropy of $\lambda$ (we write $a_N \asymp
b_N$ when $\ln{a_N}/\ln{b_N} \to {\rm 1}$ as $N\to\infty$). We
introduce the {\em entropic complexity}\/ $\Sigma_s(s)$ that counts the number
$\mathcal{N}_N(s)\asymp\exp[N\Sigma_s(s)]$ of clusters with internal entropy
$N s$, and a method for computing $\Sigma_s(s)$  and the total entropy
density $s_{\rm tot}$, yielding the total number of solutions, $|\mathcal{S}| \asymp
\exp(N s_{\rm tot})$, for individual
instances of any CSP. The problem of
counting the number of solutions of a CSP is in general
$\#P$-complete~\cite{Valiant79,Papadimitriou94}, a class of problems even
harder than NP-complete~\cite{Toda89}. Estimating $|\mathcal{S}|$ is important
in applications such as graph reliability~\cite{Valiant79}, 
and computing partition functions. 

A second, related aspect of the structure of $\mathcal{S}$ is its geometry. We
introduce a method to compute the {\em geo\-metric complexity}\/ $\Sigma_d(d)$,
which counts the number of clusters at a given distance $N d$ from a reference
assignment (see Fig.~\ref{fig:scheme}a), and the related {\em weight
enumerator function}\/, of direct interest in coding theory~\cite{MacKay03}.
Finally, we indicate several generalizations of these methods.

Our methods are based on extensions of the ``energetic'' cavity method (CM) of
Ref.~\onlinecite{MezardParisi03}. We illustrate them for $q$-COL and show
numerical results for $q=3$, but we emphasize that any CSP can be studied
along the same lines.
The energy function associated to $q$-COL is that of the antiferromagnetic
Potts model, $E[\sigma]=\sum_{( i,j)}\delta_{\sigma_i,\sigma_j}$, where
$\sigma_i\in\{1,\dots,q\}$ and the sum is over the $M$ graph edges. We study
Erd\H{o}s-R\'enyi random graphs \cite{Bollobas01}, constructed by connecting
any pair of nodes with probability $2\alpha / N$. For large $N$ this gives $M
= \alpha N$ and a Poisson-distributed connectivity with mean $2\alpha$.

\begin{figure}[b]
\centering \epsfig{file=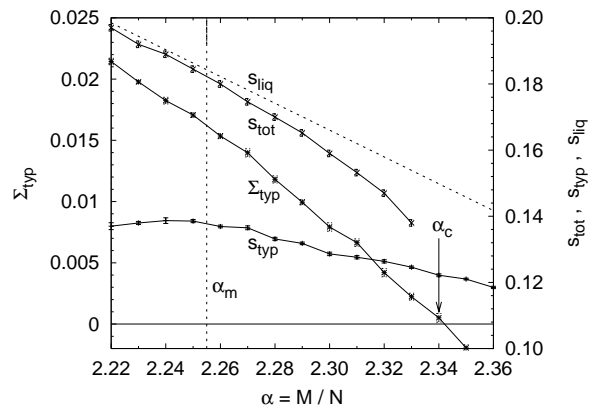,width=5.5cm,angle=-90}
\caption{\small Graph-averaged 
$\Sigma_{\rm typ}$, $s_{\rm typ}$, and $s_{\rm tot}$. Notice the different vertical
  scales. Data obtained
 with a population of $16000$ links, and $N_P=512$ fields 
on each link. 
The vertical line shows the threshold $\alpha_m \simeq
2.255$ below which the 1RSB
Ansatz is unstable~\cite{KrzakalaPagnani04}. The straight line 
is the ``liquid'' or infinite--temperature solution, 
$s_{\rm liq} = (1-\alpha) \ln(q)+\alpha \ln(q-1)$.
}\label{fig:averages}
\end{figure}

In the unclustered phase (``replica symmetric'' phase in the language of
spin-glass theory) the zero-temperature energetic CM~\cite{MezardParisi03}
computes the ground state energy recursively by adding one node at a time.
For large enough $\alpha$, 
the recursion no longer admits a unique solutions and is generalized, 
via the one-step
replica symmetry breaking Ansatz (1RSB), to a distributional recursion which
can be solved self-consistently, yielding the {\em
``energetic'' complexity}\/ $\Sigma_\epsilon(\epsilon)$, which counts the
number $\mathcal{N}_N(\epsilon)\asymp \exp[N\Sigma_\epsilon(\epsilon)]$ of
clusters of local minima with energy $E=N\epsilon$. In particular, $\Sigma_\epsilon(0)$ is
found positive in an interval $\alpha \in [\alpha_{d}^{(f)},\alpha_c]$.
The method was applied to $q$-COL in
Refs.~\onlinecite{MuletPagnani02,BraunsteinMulet03}, that report
$\alpha_d^{(f)}\simeq 2.21$, $\alpha_c\simeq 2.34$  for $q=3$ (see also
Fig.~\ref{fig:averages}). The validity of the 1RSB Ansatz in an interval
$[\alpha_m, \alpha_{SP}]$ containing $\alpha_c$ was established for $q$-COL in
Ref.~\onlinecite{KrzakalaPagnani04}, using the stability analysis of
Ref.~\onlinecite{MontanariRicci03}, with $\alpha_m \simeq 2.25$, $\alpha_{SP}
\simeq 2.50$ for $q=3$.

\paragraph*{Counting solutions --}
The energetic CM has the virtue of being simple enough, and it thus allows a
precise determination of $\alpha_c$, and the development of a powerful new
class of algorithms ({\em survey propagation}~\cite{MezardZecchina02}). This
simplicity is obtained because one focuses only on clusters in which
some of the variables are {\em frozen}, i.e. constrained to adopt a
unique color. Computing the entropy requires a more detailed
information, and thus a different formalism, as first identified in 
Ref.~\onlinecite{BiroliMonasson00} within the replica framework.
Our approach to computing entropies is illustrated in Fig.~\ref{fig:scheme}b.
The basic quantity we consider is the number $Z^{(i\to j)}_{\sigma_i}$ of solutions for the
``cavity'' graph obtained from the original graph by removing node $j$, when
the color of node $i$ is fixed to $\sigma_i$. In the unclustered phase, due to
the locally tree-like structure of large random graphs, the quantities
$Z_{\sigma_k}^{(k\to i)}$, with $k$ denoting any of the nodes connected to $i$
except $j$ (in symbols, $k \in i-j$), are independent of each other for large
$N$. Hence a recursion relation holds, $Z_{\sigma_i}^{(i\to j)}= \prod_{k\in
i-j} \sum_{\sigma_k\neq\sigma_i} Z^{(k\to i)}_{\sigma_k}$.
By defining a {\em cavity field} as the probability of having
color $\sigma$ on node $i$ in the absence of $j$, 
$\psi^{(i\to j)}_\sigma \equiv Z^{(i\to j)}_\sigma/\sum_\tau Z^{(i\to j)}_\tau$,
the recursion relation translates to
\begin{equation}\label{eq:entropy}
\psi_\sigma^{(i\to j)} = \hat{\psi}^{(i\to j)}_\sigma(\{\psi^{(k\to i)}\})
\equiv {\mathcal{Z}}^{-1} \prod_{k\in i-j} (1-\psi^{(k\to i)}_\sigma)
\end{equation}
with $\mathcal{Z}$ fixed by normalization. The ensemble of these equations on
all oriented links, known as belief propagation
equations~\cite{KschischangFrey01}, has a unique solution for $\alpha <
\alpha_d$. In general $\alpha_d\leq\alpha_d^{(f)}$ holds,
since $\alpha_d^{(f)}$ refers to the onset of clusters with frozen variables 
while at $\alpha_d$ clusters without frozen variables may also
appear. 
It is not difficult to show that the total entropy of the whole
graph is given by $N s_{\rm tot} = \sum_i\Delta S^{(i)}-\sum_{(i,j)} \Delta
S^{(i,j)}$ where, similarly to the energetic CM~\cite{MezardParisi02}, we need
to substract the link contributions $\Delta
S^{(i,j)} =\ln (1-\sum_\tau\psi_\tau^{(i\to j)}\psi_\tau^{(j\to i)})$
from the node contributions $\Delta S^{(i)}=\ln \sum_\tau
\prod_{k\in i} (1-\psi_\tau^{(k \to i)})$  to avoid
double counting. Above $\alpha_{d}$, following the 1RSB Ansatz~\cite{MezardParisi01}, we assume the
existence of many clusters.
We then compute a potential $\phi(x)$ related to the entropic complexity $\Sigma_s(s)$ through
\begin{equation}
e^{N \phi(x)} = \int_{s_{\rm min}}^{s_{\rm max}} e^{N[\Sigma_s(s)+x\, s]}ds \, ,
\label{eq:phi_sigma}
\end{equation}
where $x$ is a Lagrange multiplier which fixes the internal entropy
and $s_{\rm min}$, $s_{\rm max}$ are the points at which $\Sigma_s(s)$ vanishes.
Assuming the independence of the quantities
$Z_{\sigma_k}^{(k \to i)}$ {\em within}\/ each cluster, we introduce
probability distributions of the cavity fields $P^{(i\to j)}(\psi^{(i\to j)})$
with respect to the clusters, and generalize the cavity recursion to
\begin{equation}
\begin{split}
 P^{(i\to j)}(\psi^{(i\to j )})\propto &
\int\prod_{k\in i-j} dP^{(k\to i)} (\psi^{(k\to i)})
\mathcal{Z}(\{\psi^{(k\to i)}\})^x\\
& \delta\left(\psi^{(i\to j)}-\hat{\psi}^{(i\to j)}
(\{\psi^{(k\to i)}\}) \right)\ .
\end{split}
\label{Srecursion}
\end{equation}
 After solving  Eq.~(\ref{Srecursion}),
the potential is computed as
\begin{equation}\label{eq:phi}
\begin{split}
N \phi(x)  &  =  \sum_i\ln\int\prod_{j\in i}dP^{(k\to i)}(\psi^{(k\to i)})
e^{x \Delta S^{(i)} (\{\psi^{(k\to i)}\})}\\
& - \sum_{(i,j)}\ln\int \prod_{\substack{\
a\ = \
\\(i\to j),\\(j\to i)\
}}dP^{(a)}(\psi^{(a)})
e^{x \Delta S^{(i,j)}(\psi^{(i\to j)},\psi^{(j\to i)}) } \, ,\\
\end{split}
\end{equation} 
where $\Delta S^{(i)}$, $\Delta S^{(i,j)}$ are given above. 
A saddle point evaluation of Eq.~(\ref{eq:phi_sigma}) gives $x = - \partial_s
\Sigma_s(s)$. Hence, from $\phi(x)$ 
 we obtain $\Sigma_s(s)$ via the Legendre transform $s(x) = \partial_x
\phi(x), \Sigma_s(x) = \phi(x) - x s(x)$.
\begin{figure}
\centering \epsfig{file=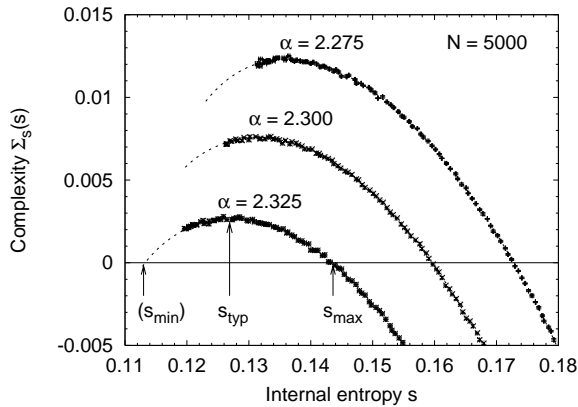,width=5.5cm,angle=-90}
\caption{\small Entropic complexity for three individual Erd\H{o}s-R\'enyi
graphs with $N=5000$ and different values of $\alpha = M/N$. Data 
obtained with a population size $N_P = 512$ on each oriented link (we verified
that using $N_P = 4096$ gives a change smaller than the error bars). 
The dotted lines are obtained by a polynomial fit
of the potential $\phi(x)$, the symbols by direct computation of the
derivative $\partial_x \phi(x)$ in Eq.~(\ref{eq:phi})
\cite{fnNum}.}\label{fig:entropy}
\end{figure}
We solve numerically Eq.~(\ref{Srecursion}) on individual
graphs by representing the distributions $P^{(i\to j)}$ with a population of
$N_P$ cavity fields on each oriented link. The resulting message passing
algorithm is an entropic generalization of survey propagation
\cite{MezardZecchina02}. Our entropic CM provides greater information 
at the price of greater computational difficulty, due to
the continuous nature of the cavity fields. 

Figure~\ref{fig:entropy} displays some of our results for 3-COL, 
for three individual graphs with $N=5000$. In particular,
the total entropy of solutions 
is given by $s_{\rm tot} = \phi(1) = s_{\rm max}$
where the last equality holds because,
according to our numerical results in Fig.~\ref{fig:entropy},
$\Sigma_s(x)$ vanishes at $x = x^* < 1$, with $s_{\rm max} = \partial_x
\phi(x=x^*)$.

Therefore,
for 3-COL the total entropy is dominated by  a subexponential 
number of giant clusters: a randomly chosen solution 
falls almost surely in one of such rare
clusters.  We also find that the fraction of 
frozen variables is finite in the  interval $[s_{\rm min}, s_{\rm max}]$. 

We also implemented a version of Eq.~(\ref{Srecursion}) averaged
over Erd\H{o}s-R\'enyi graphs, 
by considering a population of links with Poisson
connectivity and a population of cavity fields on each link. 
Figure~\ref{fig:averages} shows the graph averages obtained in this way
for $s_{\rm tot}$,
the typical  internal entropy  $s_{\rm typ} = \arg\max_s \Sigma_s(s)$,
and the typical
complexity $\Sigma_{\rm typ} = \Sigma_s(s_{\rm typ}) [ =
\Sigma_\epsilon(0) ]$, as a function of $\alpha$. 
The graph-averaged complexity curves $\Sigma_s(s)$ (not shown)
resemble those in Fig.~\ref{fig:entropy}.  
(Graph-to-graph fluctuations for $N=5000$ are significant: for
$\alpha = 2.3$, the standard deviation of $\Sigma_{\rm typ}$ is 
about $27\%$ of the mean).
The negative complexities in Fig.~\ref{fig:entropy}
have no direct interpretation on individual graphs, but for the
graph-averaged 
case they are related to rare atypical graphs~\cite{Rivoire05}.

The above formalism can be generalized to 
yield $\Sigma_{\e,s}(\e,s)$, the complexity associated with 
metastable clusters of energy
$N\e > 0$ and entropy $Ns$, with
$\Sigma_{\e,s}(0,s)=\Sigma_s(s)$,  by
adding a second  multiplier $y$~\cite{MezardPalassini06}. An equivalent
information is contained in the finite temperature complexity
$\Sigma_f(f;\beta)$~\cite{MezardParisi01}, where $f$ is the free energy and
$\beta$ the inverse temperature, based on the identity
\begin{equation}
\int  e^{N[\Sigma_{\e,s}(\e,s)-y\e+xs]}d\e\ ds=\int
e^{N[\Sigma_f(f;\beta)-x\beta f]}df,
\label{eq:finiteT}
\end{equation}
with $f=\e-s/\beta$ and $y=\beta x$.
The energetic CM is recovered for $\beta\to\infty$
and $x\to 0$ with $y=\beta x$ fixed, which amounts to
ignore all entropic effects \cite{fnEva}.

\begin{figure}
\centering \epsfig{file=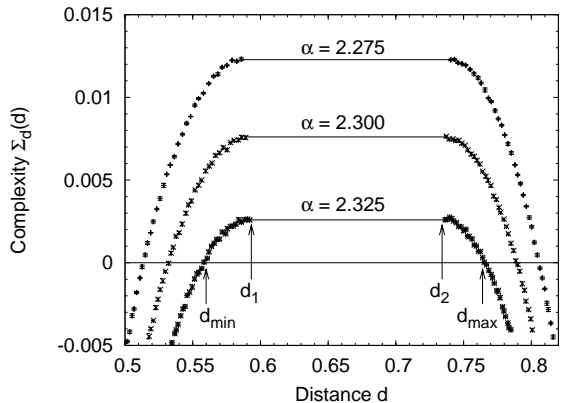,width=5.5cm,angle=-90}
\caption{\small 
Geometric complexity for the same individual graphs as in 
Fig.~\ref{fig:entropy}. Within the error bars,
$\Sigma_d(d_1)= \Sigma_d(d_2) = \Sigma_s(s_{\rm typ})$. The horizontal
lines are a guide to the eye.}\label{fig:distance}
\end{figure}


\paragraph*{Counting clusters at a given distance --}
We now turn to the geometric structure and
show how the CM can be used to investigate 
inter-cluster distances. We illustrate this
by addressing the problem of counting the number of clusters as a
function of their distance
from a fixed reference configuration $\varsigma$, which
we rephrase as a new CSP, named dCSP, whose
thermodynamics reflect the geometry of the
solution space of the initial CSP. The valid assignments of dCSP
are the solutions $\sigma \in \mathcal{S}$ of the initial CSP: 
these are
configurations of zero energy, and 
in this sense dCSP concentrates on the zero temperature case of the original problem.
But we introduce in dCSP a new energy function which is 
 the Hamming distance from $\varsigma$, $E_D[\sigma] 
\equiv\sum_{i=1}^N(1-\delta_{\varsigma_i,\sigma_i})$.
 Therefore the clusters
(resp., assignments) of dCSP with energy $E_D$ 
are the zero-energy clusters (resp., solutions) at distance $E_D$ 
from $\varsigma$ in the initial CSP.

The optimization problem for dCSP consists in finding the
maximal (or the minimal) distance between $\varsigma$ and a solution 
of the original problem.
By applying the energetic CM to this problem~\cite{fnDist},
we  obtain a {\em geometric complexity}\/ $\Sigma_d(d)$ giving the number 
of clusters at distance $N d$
of $\varsigma$, $\mathcal{N}_N(d)\asymp\exp[N\Sigma_d(d)]$.
Figure~\ref{fig:distance} shows results for 3-COL on
individual graphs. Two features are worth
noticing: {\em{i)}}  $\Sigma_d(d)$ becomes positive only above a 
threshold $d_{\rm min}$, reflecting the fact that  clusters are well separated;
 {\em{ii)}} a plateau appears between
$d_{1}$ and $d_{2}$, reflecting
the finite diameter of clusters.
We have verified that the size of this plateau coincides
with the typical diameter computed within
the entropic CM~\cite{MezardPalassini06}.

\paragraph*{Generalizations --}
The above method can be extended to count the number of {\em solutions}\/
at distance $N d$ from $\varsigma$, 
known as the weight enumerator function $A_N(d)$ in coding theory
\cite{MacKay03}.
This can be deduced from the complexity $\Sigma_{d,s}(d,s)$
which gives the number of clusters with internal entropy $N s$ at distance $N d$ from the
reference configuration $\varsigma$. Such a complexity can be obtained by
studying the dCSP with a finite value of a new inverse temperature 
$\beta_D$, which is  conjugate to the
energy $E_D$ (keeping the original temperature $\beta^{-1}$ to zero)
~\cite{MezardPalassini06}. 
Once $\Sigma_{d,s}(d,s)$
has been found, one obtains the leading
behaviour of the weight enumerator as $A_N(d) \asymp \exp[N\max_s(\Sigma_{d,s}(d,s)+s)] $.
In the same spirit, our
analysis can be extended to metastable configurations: in order to compute
the complexity
$\Sigma_{\e,d,s}(\e,d,s)$ counting clusters with energy $N \e$,
entropy  $N s$,
at distance $N d$ from $\varsigma$, one needs to introduce three
Lagrange multipliers $x,y,z$. All the previous complexities are particular limits
of this more general framework~\cite{MezardPalassini06}.


\paragraph*{Conclusions --}
We have presented methods to analyze the entropic and
geometric structure of the clustered phase in $q$-COL,
which give access to quantities 
such  as internal cluster entropies
not accessible to previous methods.
Our results for 3-COL show the existence of 
giant, atypical clusters which contain the majority of solutions.
Generalization to other CSPs such as $k$-SAT, where a similar picture
may hold, is straightforward.

Notice that
the present results were obtained within a 1RSB ansatz,
and the stability of our solution
should thus be checked (extending the method of Ref.~\onlinecite{MontanariRicci03})
to assess whether the solution is exact
or only an approximation to a more complicated one involving 
higher order RSB. 

The new information extracted with our entropic CM could be exploited 
to design new algorithms for finding solutions to individual instances,
improving on present survey propagation algorithms which only use
energetic information~\cite{MezardZecchina02}. We also 
envision applications to inference problems such as 
Bayesian belief networks~\cite{Roth96}.

We thank D.~Battaglia and R.~Zecchina for discussions, and A.~Pagnani
for sending  the SP code used in Refs.~\onlinecite{MuletPagnani02,BraunsteinMulet03}.
This work was supported in part by the European Community's Human Potential
Programme under contracts 
HPRN-CT-2002-00319 (STIPCO) and by the Community's EVERGROW
Integrated Project. 

Note added: the recent paper \cite{gamarnik}  addresses similar questions.

\bibliographystyle{apsrev}

\bibliography{reference,glasses,matchings,graphs}

\begin{thebibliography}{28}
\bibitem[{fnN()}]{fnNum}
\bibinfo{note}{The numerical solution of Eq.~(\ref{Srecursion}) becomes
  difficult for small $s$ as the distributions become broad. We only display
  data conservatively down to the lowest $s$ at which $\partial_x\phi(x)$
  computed by direct derivation of Eq.~(\ref{eq:phi}) coincides with the
  derivative of a polynomial fit of $\phi(x)$.}
\bibitem[{fnD()}]{fnDist}
\bibinfo{note}{If $\varsigma\in\mathcal{S}$, the minimal distance is zero,
  but by focusing on the metastable (glassy) states, the cavity method allows
  to ignore this trivial (crystalline) ground-state.}
\end{thebibliography}

\end{document}